\documentclass[fleqn,usenatbib]{mnras}

\usepackage{newtxtext,newtxmath}

\usepackage[T1]{fontenc}
\usepackage{ae,aecompl}

\usepackage{graphicx}	
\usepackage{amsmath}	
\usepackage{amssymb}	
\usepackage{todonotes}

\title[Flyby-induced misalignments]{Flyby-induced misalignments in planet-hosting discs}
\author[R. Nealon et al.]{
Rebecca Nealon,$^{1}$\thanks{E-mail: rebecca.nealon@leicester.ac.uk}
Nicol\'as Cuello$^{2,1,3}$
and Richard Alexander$^{1}$
\\
$^{1}$School of Physics and Astronomy, University of Leicester, University Road, Leicester LE1 7RH, UK,\\
$^{2}$Instituto de Astrof\'isica, Pontificia Universidad Cat\'olica de Chile, Santiago, Chile,\\
$^{3}$N\'ucleo Milenio de Formaci\'on Planetaria (NPF), Chile.\\
}

\date{Accepted XXX. Received YYY; in original form ZZZ}

\pubyear{2019}

\begin{document}
\label{firstpage}
\pagerange{\pageref{firstpage}--\pageref{lastpage}}
\maketitle

\begin{abstract}
We now have several observational examples of misaligned broken protoplanetary discs, where the disc inner regions are strongly misaligned with respect to the outer disc. Current models suggest that this disc structure can be generated with an internal misaligned companion (stellar or planetary), but the occurrence rate of these currently unobserved companions remains unknown. Here we explore whether a strong misalignment between the inner and outer disc can be formed without such a companion. We consider a disc that has an existing gap --- essentially separating the disc into two regions --- and use a flyby to disturb the discs, leading to a misalignment. Despite considering the most optimistic parameters for this scenario, we find maximum misalignments between the inner and outer disc of $\sim 45^{\circ}$ and that these misalignments are short-lived. We thus conclude that the currently observed misaligned discs must harbour internal, misaligned companions.
\end{abstract}




\section{Introduction}
\label{section:intro}
Detailed observations of protoplanetary discs have identified numerous substructures including rings, gaps, spirals, misalignments and warps. Mechanisms that have been proposed to generate these features range from internal disc mechanisms \citep[e.g.][]{Okuzumi:2016by,Stammler:2017bw,Gonzalez:2017bu} to planets sculpting the disc structure \citep[e.g.][]{Dipierro:2015od} as well as external companions \citep[e.g.][]{Dong:2016oc}. Rings and gaps in particular appear to be rather common features, with most of the discs in the selectively chosen bright {\sc dsharp} sample showing clearly resolved multiples of both \citep{Andrews:2018pa}. Some of these gaps can be as wide as $40$~au, as observed in AS~209 \citep{Zhang:2018ud}.

A small subset of these gapped discs additionally show significant misalignments, where the disc is best described by an inner and outer disc that do not share the same orientation. These particular systems are revealed in scattered light images, where the inner disc casts narrow, characteristic shadows on the outer disc \citep{Marino:2015rh} that are used to constrain the relative geometry tightly \citep{Min:2017oc}. Relative misalignments in broken discs range from 80$^{\circ}$ in HD 100546 \citep{Walsh:2017ic} down to 30$^{\circ}$ in DoAR 44 \citep{Casassus:2018te}, with a smattering in between \citep[e.g.][]{Benisty:2017kq,Marino:2015rh,Loomis:2017do}. Throughout this work we define moderate misalignments to be between $10^{\circ}-45^{\circ}$ and those $\gtrsim 45^{\circ}$ to be strong misalignments.

These strongly misaligned discs are currently proposed to be generated either by an internal misaligned companion \citep[e.g.][]{facchini_2013,Zhu:2018vf} or during their chaotic formation \citep{Bate:2010nh,Bate:2018ls,Sakai:2019bt}. In the former case, the disc is separated in to an inner and outer disc by either `disc breaking' \citep{nixon_2013,facchini_2013} or by the companion carving a gap \citep{Zhu:2018vf}. The misalignment of the inner companion then promotes differential precession between the inner and outer disc, leading to a range of misalignments \citep{Facchini:2017of}. These companions would likely have to be massive to drive even moderate misalignments, with a mass ratio to the host star $\gtrsim0.001$ \citep{Nealon:2018ic,Zhu:2018vf}. Indeed, numerical simulations of HD~142527 using the orbital parameters of the observed stellar companion has shown comprehensive agreement with observed disc features \citep{Price:2018pf}. Aside from HD~142527, no such companions have yet been detected \citep[although some do show suggestive features, e.g.][]{Perez:2018bh}. The long term evolution of these models has not yet been addressed.

The above scenario hinges on a misaligned companion providing a misaligned torque and on some mechanism to separate the disc into two smaller discs. However, such a torque could easily be provided by a stellar companion external to the disc, which can either be bound (binary) or unbound (flyby). In this work, we consider whether a flyby encounter may be able to provide such a torque, potentially forming strongly misaligned discs. Here, we define a flyby as the interaction between two stars at a pericentre distance of less than 1000~au.

Stellar encounters are particularly relevant in stellar clusters during the first Myr of evolution \citep{Bate:2018ls}.  Recently, \citet{Winter:2018jw} estimated that the probability of a solar-type star having such an encounter is of 20 per cent or more after 3~Myr, considering various stellar densities within clusters. However, the rate of encounters quickly decreases over time \citep{Pfalzner:2013bu}. It is precisely when flybys are the most likely that the probability of having circumstellar discs is the highest \citep{Scally:2001bh,Williams:2011he}. Therefore, flybys are expected to alter disc evolution.

There have been numerous numerical studies investigating the evolution of a disc in response to a flyby encounter. Warping of the primary disc has been found at a range of encounter inclinations and closest approach distances \citep[e.g.][]{Clarke:1993bv,Ostriker:1994ph,Terquem:1996qh,Bhandare:2016kw,Xiang-Gruess:2016bq} as well as when considering a bound companion \citep{lubow_ogilvie_2000,Lubow:2001qk}. As the pericentre passage occurs, mass in the disc tends to be redistributed and moves inwards as the mass accretion rate onto the primary star increases \citep{Ostriker:1994ph,Pfalzner:2003jh,Pfalzner:2004qi,Moeckel:2006dp,Forgan:2009ao}. Tidal effects by the perturber also leads to formation of spirals and can promote fragmentation in the disc \citep{Ostriker:1994ph,Pfalzner:2003jh,Shen:2010nk,Thies:2010wj} but when radiation effects are included, fragmentation is discouraged \citep{lodato_pringle_2007,Forgan:2009ao}. Most recently, \citet{Cuello:2019bd} simulated both the gas and dust disc during an encounter. While confirming previous results, they established that the evolution of the more compact dust disc can also reveal signatures of the flyby \citep{Cuello:2019me}. While warping of the disc is well established, none of the previous work has shown that a flyby is capable of breaking a disc.

In this work, we thus focus on the combination of a flyby and a disc that has an existing gap --- potentially driven by a massive planet or multiple planets. In this scenario the tidal torque from the planet opens a gap, efficiently separating the disc with a morphology that depends on planet mass, star mass, disc aspect ratio and viscosity \citep[e.g.][]{Kanagawa:2015if}. The evidence from the recent {\sc dsharp} \citep{Andrews:2018pa} Taurus-Auriga \citep{Long:2018vt} surveys strongly support the scenario of (sub-)Jupiter mass planets carving wide gaps in the disc at large distances from the star \citep[also][]{Lodato:2019iw}. Such discs exposed to a flyby will be subjected to differential torques, possibly driving a relative misalignment between the discs.

The flyby may also drastically affect the planet orbit by either capturing the planet \citep[e.g.][]{Breslau:2019qr} or increasing both the eccentricity and inclination of the planet orbit \citep{Hao:2013kb}. In the latter case, with a hyperbolic flyby and multiple discs, \citet{Marzari:2013iq} showed that these features are damped within about 10~kyr due to interactions between discs. Three-dimensional simulations suggest that while moderate misalignments can be obtained between the planet and disc \citep{Picogna:2014if}, large misalignments or eccentricities are not likely to be driven. These studies suggest that planet-disc interactions rapidly removes any evidence of the flyby \citep[even when the disc mass is low, as in][]{Marzari:2013iq}.

In this work we use numerical simulations to investigate whether a broken disc can be strongly misaligned by a stellar flyby. In Section~\ref{section:method} we describe our numerical simulations and the associated initial conditions. In Section~\ref{section:results} we present the evolution of the discs subjected to flybys and examine the relative tilt that develops between the inner and outer disc as well as the evolution of the planet. Section~\ref{section:discussion} expands on the observational consequences of our findings and we conclude in Section~\ref{section:conclusion}.


\section{Numerical method}
\label{section:method}

We use the SPH code \textsc{Phantom} to conduct these simulations \citep{Phantom}. A Lagrangian method is preferred for these simulations as there is no restriction on the geometry of the disc --- in the case of a flyby encounter, the disruption of the disc means its geometry may be significantly altered. \textsc{Phantom} is particularly well suited to simulations of warped discs \citep[e.g.][]{lodato_2010,Nealon:2016lr}, flybys \citep[e.g.][]{Cuello:2019bd} and planet-disc interactions \citep[e.g.][]{Dipierro:2015od,Nealon:2018ic}. In this work we consider only gas simulations, noting that the observational signatures of misaligned discs have been thoroughly explored  \citep[e.g.][]{Juhasz:2017rn,Facchini:2017of,Price:2018pf,Nealon:2019bw}.

We conduct a suite of simulations to investigate whether stellar flybys can drive appreciable misalignments in protoplanetary discs. Described below, our numerical setup begins with a disc that has a planet on an orbit that is coincident with the disc mid-plane. We assume this planet is able to carve a gap in the disc at its orbital radius, potentially separating it into an inner and outer disc. The parameters of the subsequent flyby are chosen to maximise the warping of the disc whilst minimising destruction of the outer material. Throughout the paper our results are presented in units of the orbital period at the initial planet radius of 15~au.

\subsection{Disc and planet initial conditions}
Each disc  extends from $R_{\rm in}=1$~au to $R_{\rm out}=50$~au, where $R$ is the cylindrical radius measured from the primary star. The disc has a total mass of $0.01\,M_{\odot}$ and we do not consider self-gravity effects. In these simulations the planet and stars are both affected by the presence of the gas (i.e. the planet is able to migrate) but because of the low disc mass we choose to neglect disc self-gravity.

The gas in the disc is modelled as locally isothermal with a sound speed $c_{\rm s}(R) \propto R^{-q}$. Here $q=14/$ such that the temperature in the disc, $T \propto R^{-1/2}$ \citep{Kenyon:1987bg,Andrews:2007pr,Williams:2011he}. The disc thickness is set by the aspect ratio, with  $H/R = 0.05$ at 1~au. In each simulation the gas disc is set up with its angular momentum parallel to the $z$ axis (that is, the disc lies in the $x-y$ plane).

For consistency across our different parameter choices, we chose to initialise each disc with an existing gap around the planet orbit. As there is no delay for the planet to carve the gap, this initial condition ensures that the discs in all of our simulations have evolved the same amount before the flyby occurs. The `pre-carved' gap is set to be $\Sigma_{\rm gap}$ deep and has a width of twice the Hill radius centred on the planet orbit ($\pm 2R_{\rm H}$). That is, we set the surface density profile in the disc such that
\begin{equation}
    \Sigma(R) = \begin{cases}
    \Sigma_0 \left(\frac{R}{R_0}\right)^{-p} \left( 1 - \sqrt{\frac{R_{\rm in}}{R}}\right) &|R - R_{\rm p}| > R_{\rm H}\\
    \Sigma_{\rm gap} & |R - R_{\rm p}| =< R_{\rm H},
    \end{cases}
    \label{equation:sigma}
\end{equation}
with $\Sigma_0$ determined from the total disc mass and $p = 1$ \citep{Pringle:1981fo}. Here $\Sigma_{\rm gap}$ is initially set by the approximation provided in Equation 41 of \cite{Kanagawa:2015if}.

\begin{figure}
    \centering
    \includegraphics[width=\columnwidth]{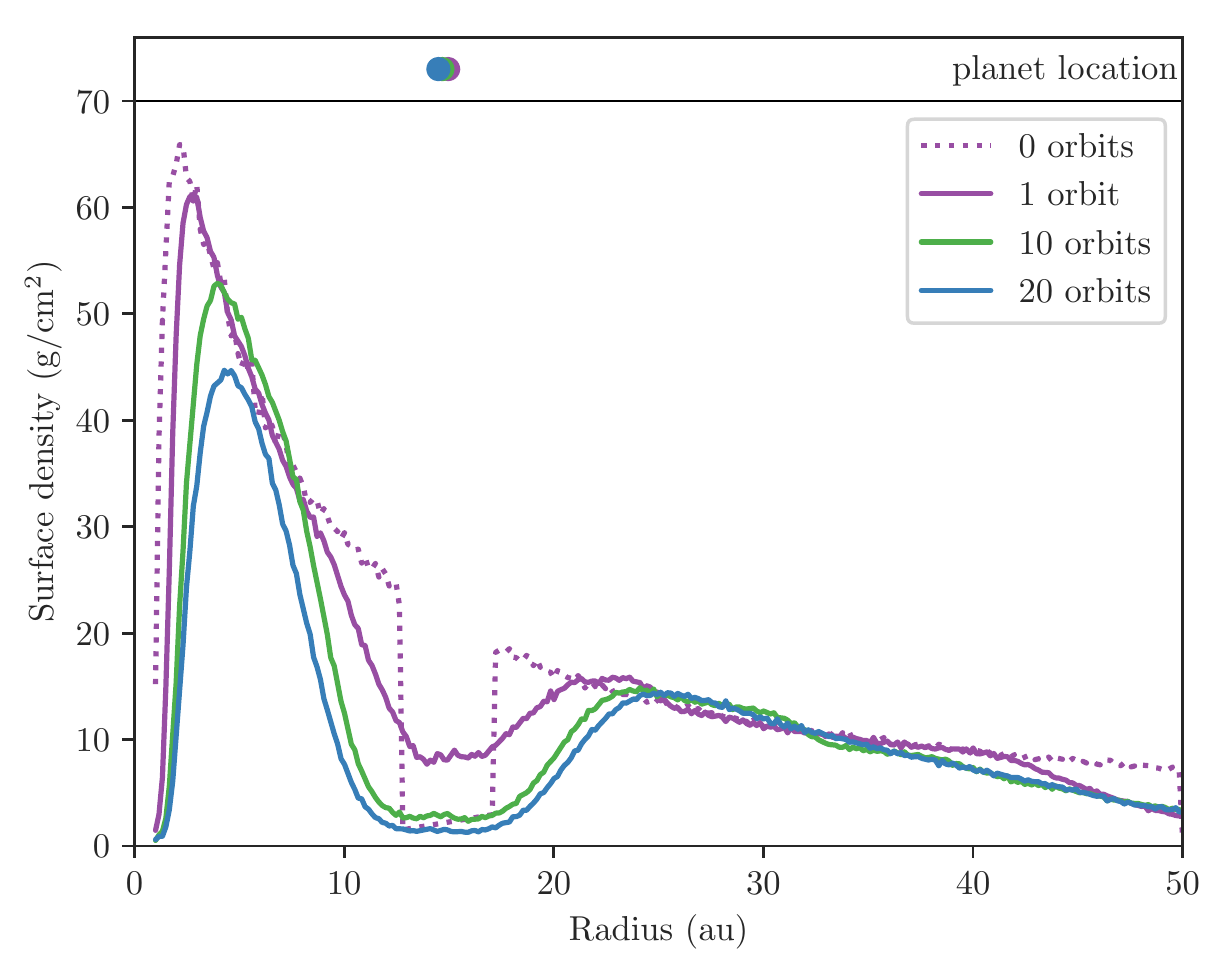}
    \caption{Evolution of the surface density profile before pericentre passage occur for our disc with a $10\,M_{\rm J}$ planet. The disc is initialised with a gap at the planet radius, which quickly smooths to a consistent profile. The transients due to this initial condition rapidly die down before the perturber arrives (at 30 orbits). The planet location is indicated by the filled circles in the upper panel, showing that there is little migration during this time.}
    \label{fig:sigma_IC}
\end{figure}

Figure~\ref{fig:sigma_IC} shows the development of the surface density profile before pericentre passage. The evolution in the first ten orbits shows that the initial gap profile we choose is not important as it is smoothed out rapidly, with the transients due to the initial conditions dissipated after approximately 20 orbits. In Table~\ref{tab:summary_table} we estimate the actual gap profile just before pericentre passage using the method outlined in \citet{Zhang:2018ud}.

The primary star, planet, and perturber are all modelled using sink particles \citep{Phantom}. Unless otherwise stated, we use $R_{\rm p} = 15$~au and vary the planet mass between 1, 5 and $10\,M_{\rm J}$. The planet accretion radius is set to $1/8$ of the Hill radius \citep[see appendix of][]{Nealon:2018ic}. The primary star has a mass of $M_* = 1\,M_{\odot}$ and both stars have an accretion radius of $1$~au.

To correctly capture shocks, particle methods include an artificial bulk viscosity coefficient denoted by $\alpha_{\rm AV}$ \citep{Lucy:1977lr,gingold_monaghan_1977}. This artificial viscosity can be related to the kinematic viscosity (i.e. $\nu = \alpha c_s \Omega$, where $\Omega$ is the Keplerian angular velocity) via \citep{Murray:1996po,lodato_2010}
\begin{align}
\alpha= \frac{\alpha_{\rm AV}}{10} \frac{ \langle h \rangle}{H}\,\,\,.
\end{align}
Here $\langle h \rangle$ is the azimuthally averaged smoothing length and $\alpha$ is the \citet{shakura_sunyaev} viscosity parameter. In our simulations, $\alpha_{\rm AV}$ is set to give an $\alpha = 0.001$ for a given $\langle h \rangle/H$. With $N=1 \times 10^6$ particles between $R_{\rm in}$ and $R_{\rm out}$, this corresponds to $\langle h \rangle /H < 0.5$ outside of 5~au during pericentre passage.

\subsection{Initialising the fly-by}
In this work we chose the parameters of our flyby to maximise warping (and potential misaligning) of the disc whilst minimising destruction of the outer disc. We thus choose a retrograde flyby that has $r_{\rm peri} = 75$~au --- i.e. a non-penetrating encounter as $r_{\rm peri} > R_{\rm out}$ \citep{Clarke:1993bv,Xiang-Gruess:2016fj,Cuello:2019bd}. We assume the perturber approaches on an orbit misaligned by $135^{\circ}$, as \citet{Xiang-Gruess:2016fj} has shown that this particular angle produces the largest misalignment of a single disc. Finally, we choose three perturber masses of $M_{\rm pert} = 1$, $5$ and $10\,M_{\odot}$ \citep{Pfalzner:2013bu}. We comment in Section~\ref{subsection:implications} on the relative likelihood of each of these perturber masses.

We initialise the perturber at a distance such that the time to pericentre passage $t_{\rm peri} = 30$ planet orbits. This choice allows both the the transients from the initial condition to settle ($\sim$20 orbits from Figure~\ref{fig:sigma_IC}) and extra time to account for the perturber effecting the disc just before and after pericentre passage. The properties of the flyby are setup as described in Appendix~A of \citet{Cuello:2019bd}. All simulations are followed for at least 80 planet orbits at $R_{\rm p}=15$~au, where the perturber is no longer affecting the disc dynamics.

\begin{table}
	\centering
	\caption{Table summarising the parameters used for our main simulations. Here $m_{\rm p}$ is the planet mass, $\Delta R_{\rm gap}$ is the width of the gap and $\delta \Sigma$ is the gap depth at pericentre passage \citep[both calculated following][]{Zhang:2018ud} and $M_{\rm pert}$ is the perturber mass.}
	\label{tab:summary_table}
	\begin{tabular}{lcccr} 
   Name & $m_{\rm p}$ ($M_{\rm J}$) & $\Delta R_{\rm gap}$ (au) & $\delta \Sigma$ & $M_{\rm pert}$ ($M_{\odot}$) \\ 
		\hline
		R1 & 1.00 & 4.92 & 0.58 & 1.00\\
		R2 & 1.00 & 4.91 & 0.58 & 5.00\\
		R3 & 1.00 & 3.81 & 0.65 & 10.0\\
		R4 & 5.00 & 10.7 & 0.11 & 1.00\\
		R5 & 5.00 & 9.52 & 0.13 & 5.00\\
		R6 & 5.00 & 8.36 & 0.16 & 10.0\\
		R7 & 10.0 & 13.1 & 0.04 & 1.00\\
        R8 & 10.0 & 12.5 & 0.04 & 5.00\\
        R9 & 10.0 & 11.2 & 0.07 & 10.0\\
        R10 & 1.0 & 22.1 & 0.20 & 10.0\\
		\hline
	\end{tabular}
\end{table}


\subsection{Measuring the evolution of the disc}
The evolution of the disc is characterised by the direction of the disc angular momentum vector. This is described with the two angles of tilt $\beta$ and twist $\gamma$ such that the unit angular momentum vector of the disc is \citep{pringle_1996}
\begin{align}
    \boldsymbol{\ell}(R,t) = (\cos \gamma \sin \beta, \sin \gamma \sin \beta, \cos \beta)\,\,\,,
    \label{equation:definitionofell}
\end{align}
and the total angular momentum vector of the disc is denoted by $\mathbf{L}_{\rm disc}(R,t)$. As the disc evolves, we  define the disc material to be any gas that is bound to the primary star within 150~au (but increasing this outer boundary does not affect our results). In this work we initialise the disc with $\mathbf{L}_{\rm disc}$ parallel to the $z$ axis, using this axis as the reference to measure $\beta$ and $\gamma$ from.

We measure Equation~\ref{equation:definitionofell} by discretising the disc into radial annuli and averaging the particle properties around each annulus \citep[as described in][]{lodato_2010}. In contrast to previous work, here the annuli are defined by the semi-major axis of the gas with respect to the primary star, allowing for consistency as the disc becomes eccentric during the flyby encounter. To estimate the orientation of each disc, we then conduct a mass-weighted average across the radial bins:
\begin{align}
    \boldsymbol{\ell}_{\rm disc} = \frac{\Sigma \boldsymbol{\ell}_i m_i}{\Sigma m_i}\,\,\,,
    \label{equation:massweighting}
\end{align}
where $\boldsymbol{\ell}_i$ is the unit angular momentum and $m_i$ is the mass contained within each annulus. In the case of the inner disc ($\boldsymbol{\ell}_{\rm inner}$) Equation~\ref{equation:massweighting} is summed from the primary star to the planet orbit. Similarly, for the outer disc ($\boldsymbol{\ell}_{\rm outer}$) it is summed from the planet orbit to 150~au (but only including gas bound to the primary star).

The evolution of the disc or planet is measured by considering the tilt and twist throughout the encounter, using the initial angular momentum vector of the gas disc as a reference vector (i.e. the $z$-axis in our simulations). The tilt and twist are simply given by
\begin{align}
    \beta(t) = \arccos{(\boldsymbol\ell(t)})\,\,\,, \gamma(t) = \arccos{(\boldsymbol\ell(t)})\,\,\,.
    \label{equation:abs_tilt_and_twist}
\end{align}
The `effective misalignment' between the discs measured relative to the outer disc is
\begin{align}
    \Delta \beta_{\rm disc}(t) = \arccos{(\boldsymbol\ell_{\rm inner}(t) \cdot \boldsymbol\ell_{\rm outer}(t))}\,\,\,.
    \label{equation:rel_tilt}
\end{align}



\begin{figure*}
    \centering
    \includegraphics[width=\textwidth]{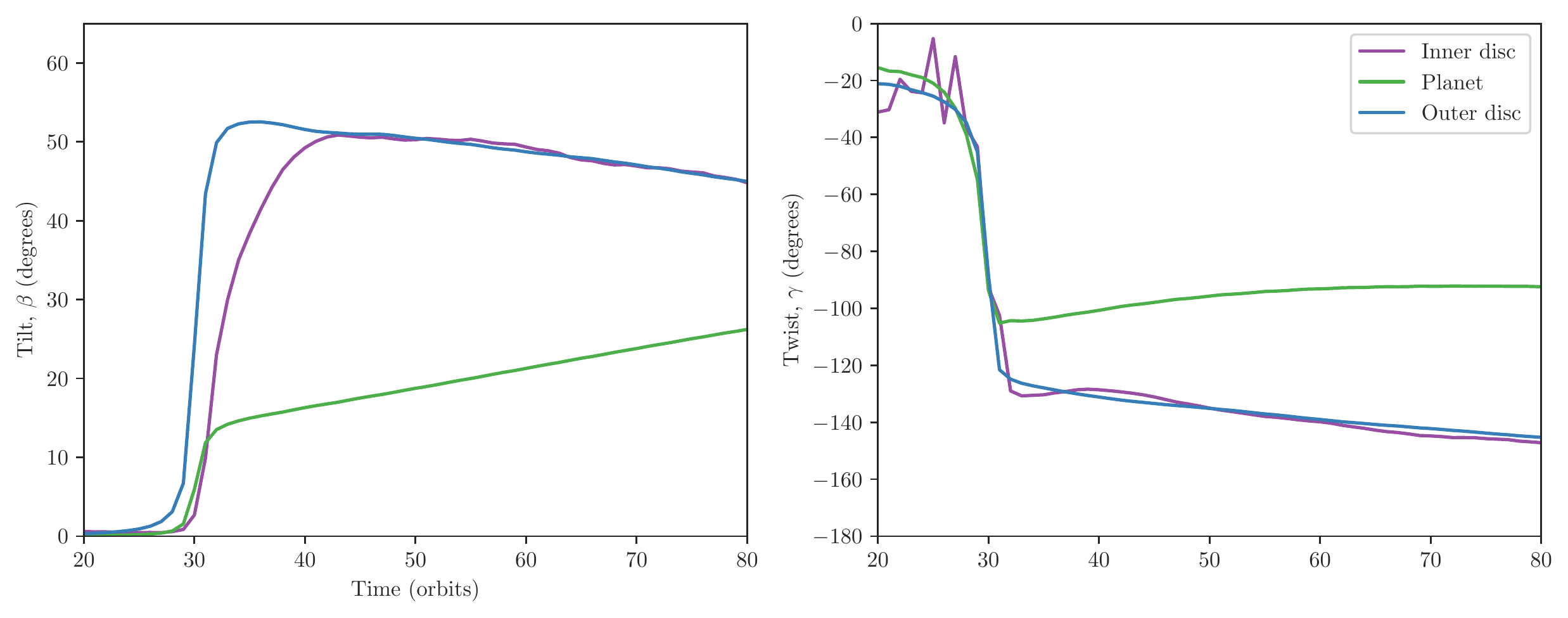}
    \caption{The relative misalignment that is generated between the inner and outer disc due to a misaligned, retrograde flyby for our representative simulation R9. Pericentre passage occurs at 30 planet orbits, measured at 15~au. Most of the relative misalignment is driven by a difference in tilt rather than twist.}
    \label{fig:misalignments_w_time}
\end{figure*}

\section{Results}
\label{section:results}
\subsection{Evolution of the disc and planet}
\label{section:withplanets}

We find broadly the same evolution occurs for the simulations R1-R9 in Table~\ref{tab:summary_table} and present the evolution of R9 in Figure~\ref{fig:misalignments_w_time}. Roughly 10 planet orbits prior to pericentre passage the discs feel the gravitational influence of the perturber and the outermost edge starts warping. During pericentre passage, the whole disc and planet are disturbed from their initial position. Approximately 15 planet orbits after pericentre passage, the disc no longer feels the influence of the perturber and the subsequent evolution is governed by viscous and pressure effects. Evolution beyond about 80 orbits is hampered by poor resolution of the inner disc and so we do not consider this here.

\subsubsection{Motion of the disc}
The disturbance from the perturber affects the outer edge of the disc first, driving a warp to the inner disc regions. Even though we have imposed a gap in each disc prior to the flyby, the encounter disrupts the disc enough that the gap is partially filled in. Once the warp is imposed at the outer edge, it propagates across the whole disc and dissipates within $\lesssim$20 orbits, erasing any obvious warp signature. A small relative inclination between the inner and outer disc mainly due to the twist does remain.

Figure~\ref{fig:misalignments_w_time} shows the evolution of the tilt and twist for simulation R9, defined by Equation~\ref{equation:abs_tilt_and_twist}. Most of the relative misalignment appears to come from a difference in tilt rather than twist. In all cases, the sharp spike of relative misalignment during pericentre passage corresponds to the outer disc tilting and twisting in response to the perturber. After passage, the inner disc rotates and over time catches up to the outer disc, eliminating the relative misalignment.

\begin{figure}
    \centering
    \includegraphics[width=\columnwidth]{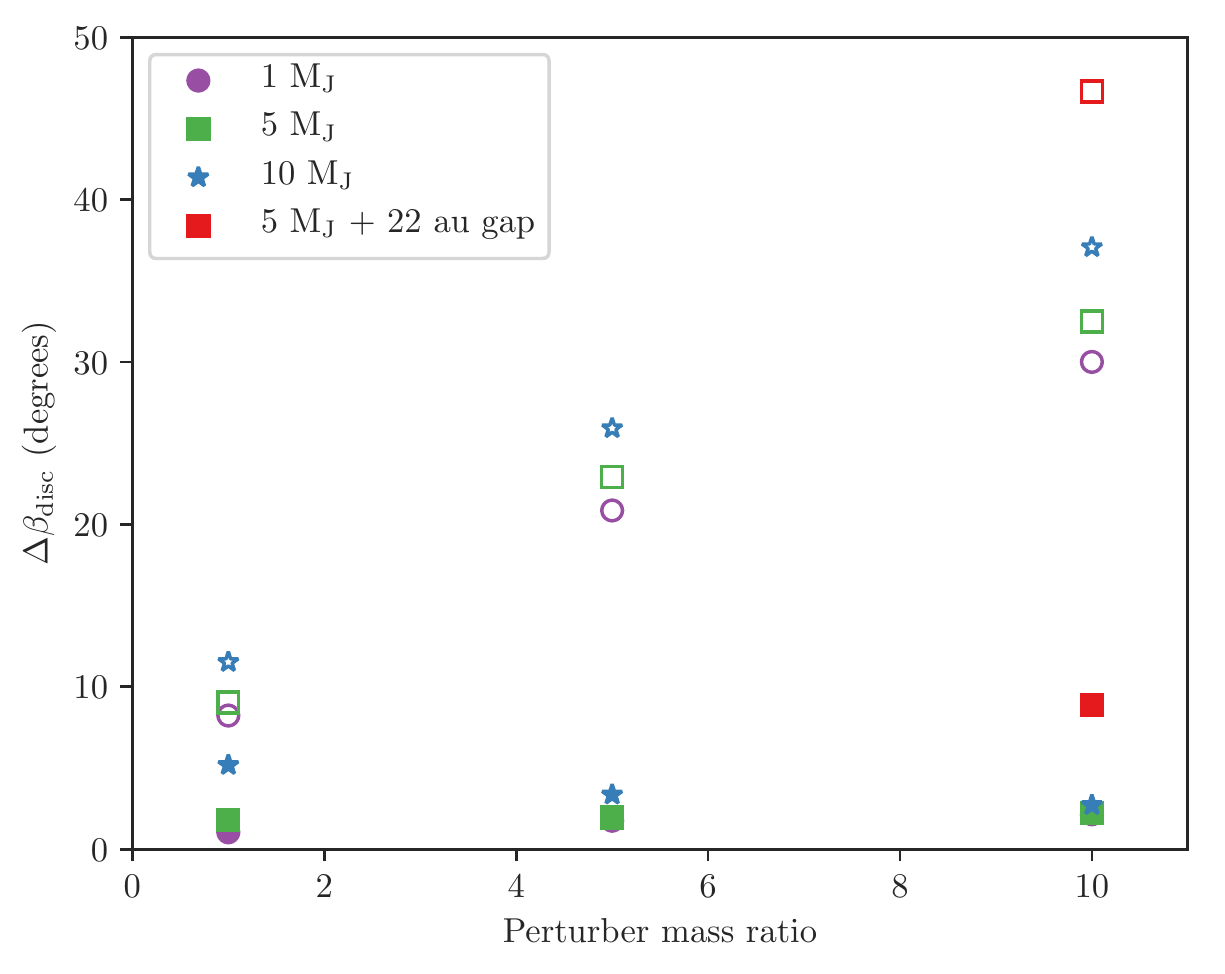}
    \caption{Maximum relative misalignment between the inner and outer disc (open symbols) and averaged over 30-50 orbits after pericentre passage (filled symbols) from all of our simulations (Table~\ref{tab:summary_table}).}
    \label{fig:rel_beta_summary}
\end{figure}

In Figure~\ref{fig:rel_beta_summary} we show the maximum and long-term relative misalignment (Equation~\ref{equation:rel_tilt}) between the inner and outer disc for all of the simulations. The maximum relative misalignment is displayed with open symbols, noting that this occurs close to $t_{\rm peri}$ as the outer disc responds to the flyby before the inner disc. Even for our most optimistic parameter choices, the discs do not develop misalignments of greater than $\sim 40^{\circ}$.

Figures~\ref{fig:misalignments_w_time} and \ref{fig:rel_beta_summary} suggest that the relative misalignment between the inner and the outer disc is greater for larger perturbers. This is due to the more massive perturbers with the same pericentre distance producing a larger disturbance to the disc and hence greater disc warping. The slight increase in the measured relative misalignment for more massive planets is likely due to the wider, deeper gap associated with the heavier planets. We test this in Section~\ref{section:biggap}. In Figure~\ref{fig:rel_beta_summary} the filled symbols indicate the average relative misalignment calculated over 60-80 orbits, well after pericentre passage and damping of the warp. The long term relative misalignment shows the opposite trend to the maximum relative misalignment, where larger sustained misalignments occur for smaller perturber masses (and for larger planet masses).


\subsubsection{Motion of the planet}
\label{section:planet_motion}

Included in our representative simulation shown in Figure~\ref{fig:misalignments_w_time} is the tilt and twist evolution of the planet. In line with the discs, the planet inclination increases rapidly during pericentre passage and then evolves more slowly due to gravitational interaction with the disc. We confirm that this subsequent evolution is due to the gravitational planet-disc interactions with the comparison shown in Figure~\ref{fig:disc_no_disc}. After the pericentre passage, the planet takes much longer than the inner disc to realign with the outer disc (Figure~\ref{fig:misalignments_w_time}). The planet-disc misalignment decreases as both align to the total angular momentum (i.e. the whole system excluding the perturber). In addition to dynamic interactions with other planets in the disc (e.g. planet-planet interactions, \citet{Nagasawa:2008pj} and resonant exchanges between planets \citet{Teyssandier:2013og,Thommes:2003is}), our results suggest that flybys may be a plausible mechanism to promote a planet onto an orbit inclined to the disc.

In line with previous work, we also identify an increase in the eccentricity of the planet orbit and a slight decrease in the semi-major axis \citep{Picogna:2014if}. In contrast to \citet{Picogna:2014if}, the planets in our simulation appear to take longer to circularise but this is due to their more highly inclined orbits. Additionally, the decrease in semi-major axis is less than noted in \citet{Picogna:2014if} because our retrograde flybys are less destructive than their prograde counterparts. In line with \citet{Fragner:2009bt} we also find the mass accretion rate onto the planet is affected by the flyby and increases for $t > t_{\rm peri}$. As expected, more massive planets are associated with larger accretion rates. Finally, we note that similar evolution is seen independent of the phase of the planets orbit during pericentre passage or for a different initial radius.

\subsection{Disc communication}

\subsubsection{Viscous and pressure effects}

To show the importance of the viscous communication on the evolution of the planet and discs, we conduct a comparison simulation to R9 that does not include hydrodynamic effects. Here we represent the inner and outer disc with sink particles that have the same mass and angular momentum as the inner and outer discs respectively. The sink particle to represent the inner disc is $2.44\,M_{\rm J}$ located at 7.5~au and the particle to represent the outer disc is $8.04\,M_{\rm J}$ located at 30~au. The sink particle for the planet remains unchanged. This setup ensures that the angular momentum balance of the system is the same but allows it to be modelled without viscous or pressure effects.

\begin{figure}
    \centering
    \includegraphics[width=\columnwidth]{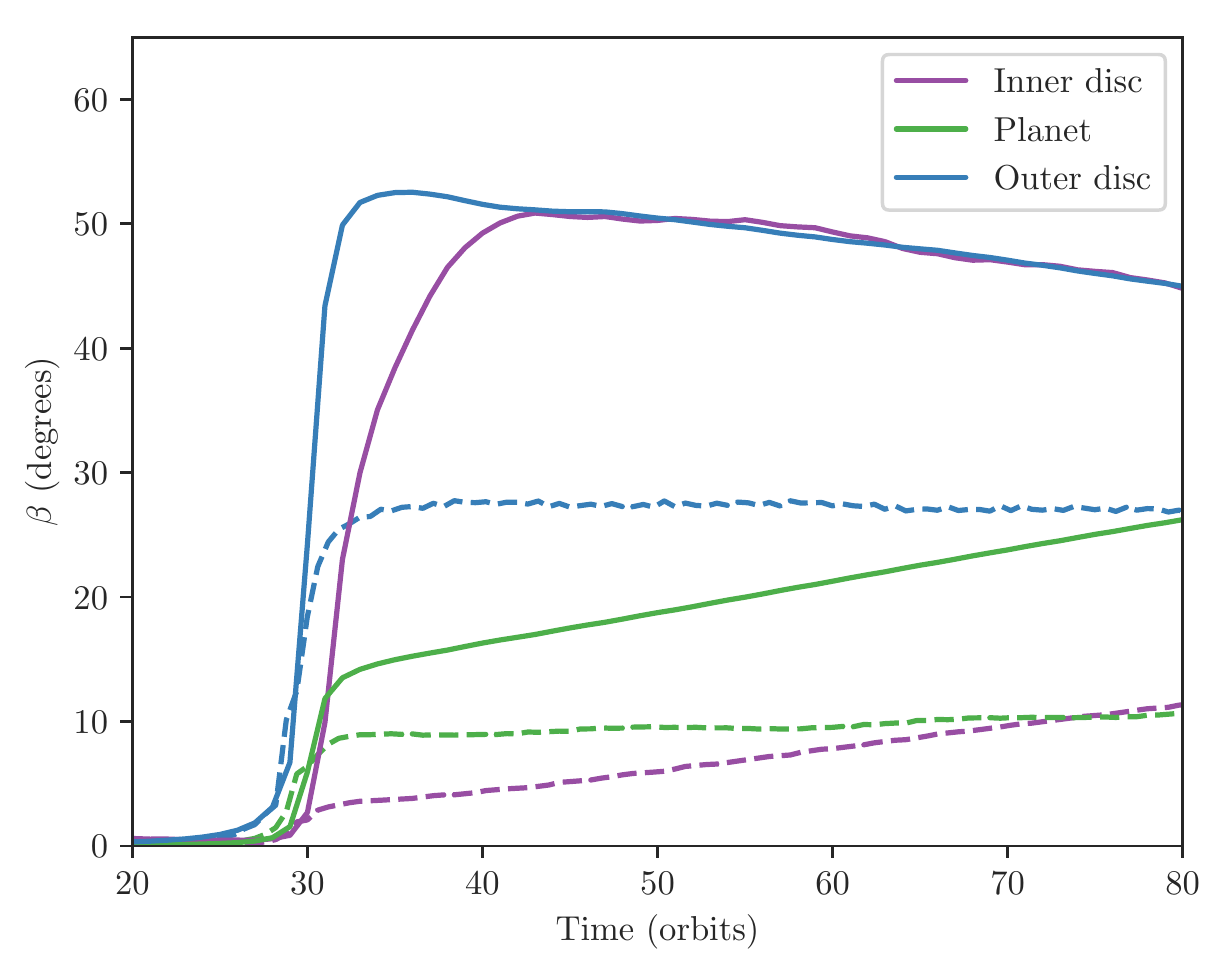}
    \caption{Effect of viscosity and pressure on the tilt of the disc and planet during the flyby. The solid lines show the R9 simulation including viscous and pressure effects while the dashed lines show a comparison simulation that does not include these effects (see text). The viscous interaction between the inner and outer disc pulls the inner disc rapidly into alignment with the outer disc.}
    \label{fig:disc_no_disc}
\end{figure}

Figure~\ref{fig:disc_no_disc} shows the evolution of the tilt for this representative simulation and that of the viscous disc case. The final tilt reached by the outer disc is clearly dependent on the combination of the gravitational interaction with the flyby and viscous behaviour of the outer disc. Similarly, the inner disc tilt is strongly governed by viscous interaction with the outer disc immediately after the flyby encounter. On longer time-scales, the planet tilt increases due to gravitational interaction with the inner and outer disc, with these components all moving towards their shared total angular momentum. We thus conclude that the viscous interaction between the two discs is the main mechanism that realigns them so quickly. This is facilitated by the material that is pushed into the gap when the disc is disturbed during the flyby passage, also noted by \citet{Picogna:2014if}, \citet{Fragner:2009bt} and \citet{Marzari:2013iq}.

\subsubsection{A disc with a wide gap}

\label{section:biggap}

The previous section showed that the inner and outer disc rapidly realigned due to viscous communication, as the gap carved by the planet is filled during the encounter because of the disruption to the disc. Here we test this with a simulation of a flyby near a disc constructed of an inner and outer disc separated by a wide, deep gap. Such a gap may be caused by either a single planet \citep[as suggested by][for AS~209]{Zhang:2018ud} or a series of smaller planets. In the case of AS~209, independent observations in scattered light \citep{Avenhaus:2018iw} and C$^{18}$O ($J=$ 2--1) emission \citep{Favre:2019wb} suggest that the surface density in the gap is approximately $1/10$ times the unperturbed gas surface density.

For the purposes of this simulation we make no assumptions about what may have caused the wide deep gap. Our disc (simulation R10) is initialised as before with a gap extending between 15~au and 35~au and to prevent material from rapidly filling the gap, a $5\,M_{\rm J}$ is placed at 31~au. The mass and location of this planet do not significantly change the subsequent evolution of the discs. We additionally rescale the disc mass to $6.25 \times 10^{-3}\,M_{\odot}$ so that it has the same $\Sigma_0$ as our previous simulations. Just before $t_{\rm peri}$ the gap depth at $R=25$~au is approximately $1/7$ times what it would be without the gap.

\begin{figure*}
    \centering
    \includegraphics[width=\textwidth]{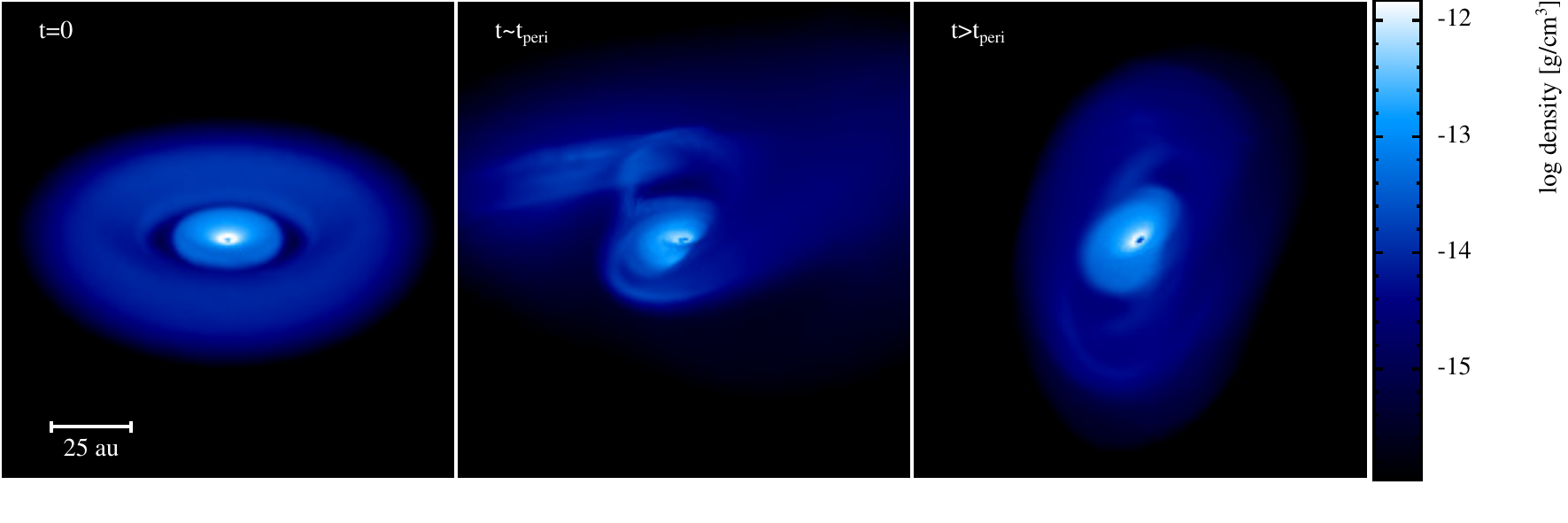}
    \caption{Density rendering of a disc with an existing, wide gap that is subjected to a retrograde flyby. The encounter causes the inner and outer discs to become strongly misaligned (by about $45^{\circ})$). This disc was initialised with a gap between 15~au and 30~au, $5\,M_{\rm J}$ at 31~au and $\eta=17.22$. The viewing orientation has been altered for clarity.}
    \label{fig:biggap}
\end{figure*}

Figure~\ref{fig:biggap} shows the rendered density structure of this disc at key points during the encounter. In line with our previous simulations, the discs achieve a strong misalignment as the flyby occurs and the gap begins to fill in after pericentre passage. As in Section~\ref{section:withplanets}, we measure the maximum misalignment achieved just after $t_{\rm peri}$ and also on longer time-scales, included in Figure~\ref{fig:rel_beta_summary} for comparison with the simulations that have a narrower gap. We find that the disc with the wider gap develops a larger misalignment ($47^{\circ}$) that is sustained for longer than in our other simulations.

The structure found in Figure~\ref{fig:biggap} confirms that large relative misalignments depend on minimal connection between the inner and outer disc. By comparing the relative misalignment generated by this wide gap simulation (R10) and simulations R7-9, we find there is a larger difference due to widening of the gap rather than changing the planet mass. This confirms that the planet does not have a significant effect on the overall disc evolution in our simulations apart from creating the initial gap. Invoking a single planet to carve a gap for reasonable planet masses (with a corresponding width listed in Table~\ref{tab:summary_table}) is much less efficient than the wide gap imposed here. However, in order to carve such a wide gap with the disc parameters chosen in our simulations we would require an object that is massive enough to be considered a stellar companion (and as such, is likely observable). When lower mass companions are observationally motivated, we thus prefer to consider multiple planets that have collectively carved a wide gap to be a more plausible explanation.

\subsection{Long term evolution}
\label{section:long_term_evolution}

Using the relative misalignment between the discs as a proxy for the alignment time-scale (Figure~\ref{fig:misalignments_w_time}), we can estimate a lower limit on how long it should take these discs to align.

We consider the 20 orbits immediately after pericentre passage for each of our simulations, measuring the rate of realignment. We fit an exponential curve and measure the damping timescale for each of our simulations with the median damping rate of $543 \pm 178$ years for our simulations with a narrow gap (the uncertainty is calculated from the standard deviation across the 9 simulations). For our simulation with a wider imposed gap we find a timescale of $577 \pm 21$ years (here the uncertainty is derived from the least squares slope fitting). Appendix~\ref{section:res_study} shows that the damping rate is consistent to within 10\% when simulated at a higher resolution.

In the case that the realignment between the inner and outer disc is governed by their viscous interaction, we can roughly estimate the rate of realignment with the viscous warp damping time-scale \citep{lubow2002evolution}
\begin{align}
    t_{\rm damp} = \frac{1}{\alpha \Omega},
    \label{equation:damping_timescale}
\end{align}
where $\alpha$ is the Shakura and Sunyaev viscosity parameter and $\Omega$ is the orbital time-scale at the warp location. For the parameters in our simulations with a narrow gap where $R_{\rm p}=15$~au and $\alpha=0.001$, the above time-scale evaluates to $\sim 9.2 \times 10^3$ years. As expected, this is longer than the crude estimate measured from our simulations. This discrepancy could be caused by the higher viscosity experienced in the low density gap carved by the planet, as we necessarily have poor resolution there (in this case, for matching time-scales we would require $\alpha \sim 0.02$ in the gap). Although this leads to faster realignment, this will not strongly affect the broader evolution of the discs as they have higher resolution far away from the gap. We thus conclude that the behaviour of the discs is robust but that alignment should take longer than represented in our simulations. Issues with viscosity in low resolution regions in SPH simulations similar to these have already been noted by \citet{Xiang-Gruess:2016bq}.


\section{Discussion}
\label{section:discussion}


\subsection{Implications for currently observed misaligned discs}
\label{subsection:implications}
Of the existing broken discs that have been observed (e.g. HD~100546, HD~135344B, AA Tau, DoAr44 and MWC758), none appear to show evidence of a warp in the outer disc that would suggest a flyby. As flybys are a rare occurrence \citep{Pfalzner:2013bu,Winter:2018jw}, this is not surprising. Additionally, the currently observed discs broadly reside in nearby clusters with low stellar densities and few high mass stars. If they have previously had an encounter, the perturber is more likely to be on the lower mass end of Figure~\ref{fig:rel_beta_summary}. This suggests a maximum relative misalignment of $\sim 10^{\circ}$ is possible, even with the optimistic parameters adopted in our simulations. We thus confirm that the unique geometry of these systems is not the result of a previous encounter with a flyby and these discs are thus very likely to host an internal misaligned companion. The exception to this discussion is of course the disc around HD~1000453A, which does show warping \citep{vanderPlas:2019gy} but this is likely due to the observed bound companion that is external to the disc \citep{Chen:2006vs,Benisty:2017kq}.

Magnetic fields may also be influencing the evolution of some of these systems. For example, the disc in AA Tau has an observed misalignment of $45^{\circ}$ \citep{Loomis:2017do} and a magnetic field that is misaligned by $20^{\circ}$ \citep{Bouvier:1999uh,Donati:2010be}. Additionally, in the case of a flyby the stellar magnetic field around the primary may act to anchor the inner disc, causing it to warp rather than freely moving as in our simulations. The inclusion of magnetic fields is beyond the scope of this work.

Remarkably, for moderate to high circumbinary disc misalignments around eccentric binaries the disc can become polar with respect to the binary orbital plane. This is due to the coupled effect of nodal libration \citep{Doolin:2011qw} and gas damping in the circumbinary disc \citep{Aly:2015tf,Martin:2017sd,Zanazzi:2018qw,Cuello:2019pb}. It is worth noting that the first polar circumbinary disc has been recently reported by \cite{Kennedy:2019qw}.


\subsection{Effect on dust dynamics and growth}
\label{sec:dustgrowth}

Flyby-induced warps in protoplanetary discs in gas and dust were recently studied by \citet{Cuello:2019bd}. In their figures 5 and 7, the distribution of dust particles with sizes ranging from $1\,\mu m$ and $10$~cm disc exhibit a comparable warp as the one observed in the gas. Therefore, the dust content of the disc is expect to closely follow the gaseous gaps in our simulations presented here. Their results showed that the main difference resides in the radial extent of the large and marginally coupled grains, which is smaller compared to the gas because of radial drift. This coupling regime corresponds to solids with Stokes numbers close to $1$. For typical protoplanetary disc parameters, this occurs for grain with sizes comprised between $0.1$~mm and $1$~cm \citep{Laibe:2012bh}. In the context of the structures formed in our simulations, we thus expect the mm-sized dust particles to be more strongly affected by the misalignment of the inner regions of the disc. This is in agreement with the \textit{in situ} formation of misaligned planets close to the star.

However, the process of grain growth within the disc strongly depends on the relative velocity among solids ($\Delta v$) and the eccentricity excitation ($\Delta e$). Low values of both $\Delta v$ and $\Delta e$ promote favourable conditions for planetesimal formation \citep{PP6_book,Blum:2018id}. As mentioned previously, retrograde encounters are less destructive and hence lead to low eccentricity excitations --- as opposed to prograde ones. Interestingly, it is possible to derive a boundary radius of the planet-forming regions around a protoplanetary disc perturbed by a stellar flyby \citep{Kobayashi:2001op}. Beyond that distance, planetesimals are expected to be destroyed during (or shortly after) the encounter. This renders \textit{in situ} planet formation in the outer regions unlikely after a flyby.


\section{Conclusions}
\label{section:conclusion}
In this work we consider whether a flyby encounter can be used to form a strongly misaligned disc. Our simulations feature an aligned planet in the disc that carves a cap, separating the disc into an inner and outer disc before the flyby occurs. The disruption by the flyby causes the inner and outer disc to move differentially, leading to a relative misalignment. We measure the relative misalignment that develops between the inner and outer disc, the warp profile that forms, the duration of these structures and the evolution of the planet orbit.

With the most optimistic parameters for the disc and flyby, our simulations show a maximum misalignment of $\sim45^{\circ}$ and that misalignments are quite short lived. During the flyby encounter, the disc is strongly disturbed and material flows into the gap carved by the planet. This allows the inner and outer disc to viscously realign rapidly. Our results thus suggest that an external perturber is not a robust method to make a strongly misaligned disc, even with parameters that are chosen to maximise the misalignment generated. Although they are yet to be directly observed, this adds to existing literature \citep[e.g.][]{Zhu:2018vf} suggesting that all currently observed strongly misaligned discs must be harbouring an internal misaligned companion.

\section*{Acknowledgements}
The authors warmly thank Giovanni Dipierro, Daniel Price, Jim Pringle and Chris Nixon for useful discussions. We additionally thank the referee for their comments that have improved the paper and in particular for the discussion that lead to Figure~\ref{fig:disc_no_disc}. NC acknowledges financial support provided by FONDECYT grant 3170680 and from CONICYT project Basal AFB-170002. This project has received funding from the European Union's Horizon 2020 research and innovation programme under the Marie Sk\l{}odowska-Curie grant agreement No 823823 (DUSTBUSTERS). This project has received funding from the European Research Council (ERC) under the European Union's Horizon 2020 research and innovation programme (grant agreement No 681601). This work was performed using the DiRAC Data Intensive service at Leicester, operated by the University of Leicester IT Services, which forms part of the STFC DiRAC HPC Facility (www.dirac.ac.uk). The equipment was funded by BEIS capital funding via STFC capital grants ST/K000373/1 and ST/R002363/1 and STFC DiRAC Operations grant ST/R001014/1. DiRAC is part of the National e-Infrastructure.




\bibliographystyle{mnras}
\bibliography{master} 



\appendix


\section{Resolution study}
\label{section:res_study}
\begin{figure}
    \centering
    \includegraphics[width=\columnwidth]{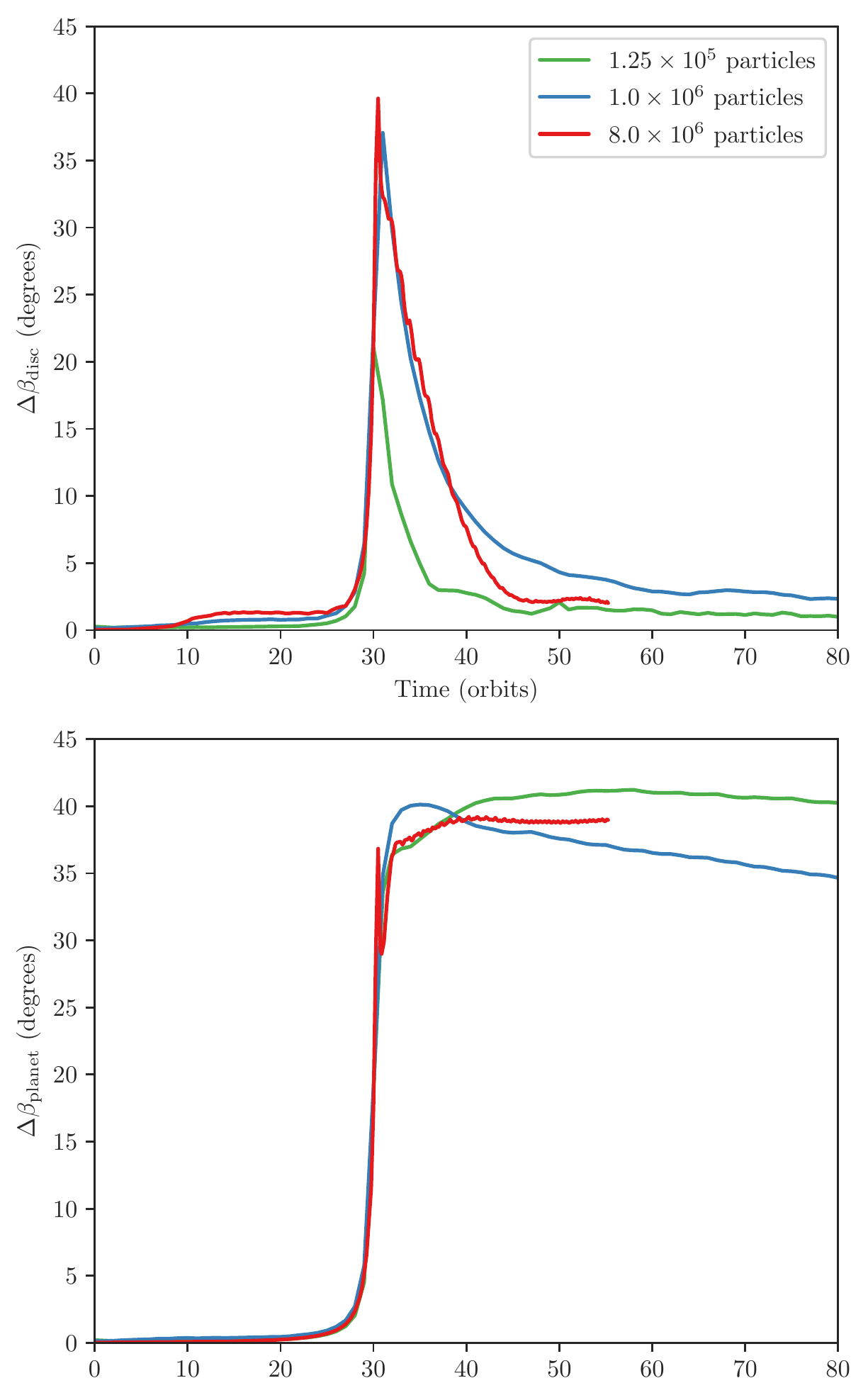}
    \caption{Movement of the planet and inner disc for a flyby with a 10~M$_{\odot}$ perturber and a 10~M$_{\rm J}$ aligned planet, conducted at three different resolutions. The upper panel shows the relative tilt between the inner and outer disc and the lower panel the tilt of the planet.}
    \label{fig:resolution_study}
\end{figure}

Here we repeat simulation R9 with $1.25 \times 10^5$ and $8.0\times 10^6$ particles (i.e. half and twice the resolution) to show that our results present converged behaviour of both the disc and planet. The upper panel of Figure~\ref{fig:resolution_study} shows the relative disc misalignment and the lower panel the relative planet misalignment for these three cases. The evolution of the relative tilt at our two highest resolutions suggests that the results in Figure~\ref{fig:rel_beta_summary} would not change significantly at higher resolution.

We repeat the estimate of the realignment rate outlined in Section~\ref{section:long_term_evolution} for each of the three resolutions. As the resolution increases, the corresponding damping time-scale is $245 \pm 23$ years, $394 \pm 20$ years and $343 \pm 4$ years. We thus conclude that the rate of damping is not strongly affected for the simulations when $N\gtrsim 10^6$ particles.

Small differences between the evolution of the discs and the motion of the planet are likely due to the resolution of the gap between the inner and outer disc, which will in turn alter the effective viscosity in the vicinity of the planet. This is reflected in the planet carved gap at pericentre passage,  which is $\sim7$ times deeper in the highest resolution case than the low resolution. However, the broad evolution of the planet is similar for all three resolutions.

\bsp	
\label{lastpage}
\end{document}